**Title:** Cerebrovascular morphology in aging and disease – imaging biomarkers for ischemic stroke and Alzheimer's disease

**Authors:** Aditi Deshpande[1], Nitya Kari[1], Jordan Elliott McKenzie[1], Bin Jiang[2], Patrik Michel[3], Nima Toosizadeh[1,4], Pouya Tahsili Fahadan[5,6], Chelsea Kidwell[7], Max Wintermark[2], Kaveh Laksari[1,8]

**Affiliations:**
[1]Department of Biomedical Engineering, University of Arizona
[2]Department of Radiology, Stanford University
[3]Department of Neurology, Centre Hospitalier Universitaire Vaudois, Lausanne, Switzerland
[4]Arizona Center on Aging, Department of Medicine, University of Arizona
[5]Neuroscience Intensive Care Unit, Medical Critical Care Service and Department of Medical Education, University of Virginia School of Medicine, Inova Fairfax Medical Campus
[6]Departments of Neurology, Johns Hopkins University School of Medicine
[7]Department of Neurology, University of Arizona
[8]Department of Aerospace and Mechanical Engineering, University of Arizona

**Abstract:**
BACKGROUND & PURPOSE: Altered brain vasculature is a key phenomenon in several neurologic disorders. This paper presents a quantitative assessment of vascular morphology in healthy and diseased adults including changes during aging and the anatomical variations in the Circle of Willis (CoW).

METHODS: We used our automatic method to segment and extract novel geometric features of the cerebral vasculature from MRA scans of 175 healthy subjects, 45 AIS, and 50 AD patients after spatial registration. This is followed by quantification and statistical analysis of vascular alterations in acute ischemic stroke (AIS) and Alzheimer's disease (AD), the biggest cerebrovascular and neurodegenerative disorders.

RESULTS: We determined that the CoW is fully formed in only 35% of healthy adults and found significantly ($p<0.05$) increased tortuosity and fractality, with increasing age and with disease – both AIS and AD. We also found significantly decreased vessel length, volume and number of branches in AIS patients. Lastly, we found that AD cerebral vessels exhibited significantly smaller diameter and more complex branching patterns, compared to age-matched healthy adults. These changes were significantly heightened ($p<0.05$) with progression of AD from early onset to moderate/severe dementia.

CONCLUSION: Altered vessel geometry in AIS patients shows that there is pathological morphology coupled with stroke. In AD due to pathological alterations in the endothelium or amyloid depositions leading to neuronal damage and hypoperfusion, vessel geometry is significantly altered even in mild/early dementia. The specific geometric features and quantitative comparisons demonstrate potential for using vascular morphology as a non-invasive imaging biomarker for neurologic disorders.

**Introduction**

Alterations in cerebral vasculature are common in the pathophysiology of many neurologic disorders. Acute ischemic stroke (AIS) and Alzheimer's disease (AD), the two most common cerebrovascular and neurodegenerative diseases, often coexist and are among the leading causes of mortality and morbidity worldwide [1]. Both conditions are hypothesized to possess significantly altered architecture of the brain vascular network [2–4]. These changes often precede clinical symptoms and may serve as early disease markers [5,6]. Consequently, studying the quantitative and qualitative changes in the cerebral vascular network due to aging and disease is pertinent to understanding brain health and may impact diagnosis, prevention, and treatment of neurologic disorders [7,8].

The reduced blood flow in AIS is coupled with a series of changes associated with structural remodeling of the cerebral vasculature that results in the vessel geometry quantitatively varying from the healthy vasculature [9,10]. Some of these changes may precede the clinical onset of the ischemic event in patients with atherosclerotic vessels [11]. Therefore, analysis of the brain vessel geometry can predict and potentially prevent AIS, improve prognostication, facilitate studying therapeutic options and guide reperfusion therapies when feasible [12]. However, quantitative studies of structural changes in the cerebral vasculature before and immediately after stroke and their effects on clinical outcomes are lacking despite their importance being stated in epidemiological studies.

Despite our expanding knowledge of the significance of cerebrovascular alterations in AD at the cellular and microvascular level [2] due to plaque and amyloid deposition as well as pathological structural alterations in the endothelium, our understanding of alterations in the major vessels architecture is still in early phases and has mainly been limited to animal studies [5,13]. Reduced cerebral blood flow (CBF) and perfusion, observed in neurodegenerative diseases, exacerbates neuronal degeneration and amyloid deposition, reduces elastin production, causes distensibility and autoregulation loss, and increases wall shear stress [2,14]. These developments could lead to significant morphological changes throughout the brain vasculature, including increased vascular stiffness, decreased diameters, and possibly higher tortuosity and fractality in AD brain vessels [2,15,16]. Therefore, stroke, atherosclerotic lesions, hypertension, and other cerebrovascular diseases (CVD) are additional risk factors for neurodegenerative disorders, in particular AD [4]. Consequently, diagnosis and research of these diseases can benefit greatly from a quantitative and qualitative analysis of the changes in the cerebral vascular network and structure in the corresponding populations [8,17].

Analysis of the pathological alteration in the cerebral vascular network requires a deep understanding of healthy vascular morphology and the normal distribution of cerebral vessels. A vascular atlas with geometric features can be used to study and quantify the variations in vascular geometry within the healthy population. As shown by our previous study, vascular morphology can be characterized using geometric properties of the vessel network, such as tortuosity, fractality (quantified using fractal dimensions), branching pattern, average diameter, total length, and volume [18]. These properties have been shown to be corroborated indicators of vascular health and potential pathology [8,18–20]. Most brain atlases typically do not include detailed morphology of the brain vascular network due to inadequate vascular imaging data from a large sample of healthy subjects [21] and a lack of validated algorithms to segment, extract, and analyze cerebral vasculature. Instead, current cerebrovascular atlases are confined to only the CoW region and not the more distal network, and lack quantification of inter-subject variations [21,22]. Developing a comprehensive atlas of the cerebral vasculature containing quantitative geometric features, including distal branches of the intracranial vessel network and normal variations at different ages in the healthy population, is essential in understanding the brain's anatomy and physiology. It also enables assessment of compensatory collateral flow in the CoW and the collateral circulation during acute and chronic ischemic conditions, given their vital role in the distribution of CBF, to assist with early diagnosis, optimal reperfusion approach, and patient outcomes prediction [11,12]. Here we present a probabilistic atlas of the healthy human cerebral vasculature, labeled based on major cerebrovascular territories, which defines the pattern of vessel distribution and quantifies vascular geometric features. We utilized this atlas to study alterations in cerebrovascular geometry and the inter-subject variations as well as vessel network modifications during natural aging. We then applied this knowledge to study vessel morphology alterations in AIS and AD, compared quantitatively to the age-matched healthy adult population to find potential hallmarks of these neurological disorders.

**Materials and Methods**

This retrospective study uses previously collected and anonymized data from IRB approved studies, made publicly available or obtained upon obtaining consent.

We used magnetic resonance angiography (MRA) scans from 175 healthy subjects, 45 AIS, and 50 AD patients for this study (Table 1). Out of the 175 healthy subjects scans, 109 were obtained from the MIDAS public database, made available by the CASILab at the University of North Carolina at Chapel Hill and distributed by Kitware, Inc. [23]. The healthy atlas excludes subjects with a past medical history of hypertension, diabetes, or head trauma. The MRA scans of the remaining 66 healthy subjects and 50 AD patients were obtained from the OASIS-3 study, conducted by the Knight Alzheimer Research Imaging Program at Washington University [24]. The AD scans correspond to patients with varying dementia levels based on the Clinical Dementia Rating (CDR), including early-mild (CDR 0.5 or 1) and moderate-severe (CDR 2 or 3) dementia. The MRA scans of the 45 AIS patients were obtained from the Centre Hospitalier Universitaire Vaudois in Lausanne, Switzerland.

**Table 1.** The imaging material for the study.

|  | N (Female) | Age (Mean ± Std. Deviation) | Modality | Resolution |
|---|---|---|---|---|
| **Healthy Subjects** | 109 (57) | 30 ± 9.3 | ToF, T1 | $0.5 \times 0.5 \times 0.5$ mm$^3$ |
|  | 66 (29) | 77 ± 9.7 |  | $0.27 \times 0.27 \times 0.27$ mm$^3$ |
| **Ischemic Stroke** | 45 (20) | 53 ± 16.1 | ToF | $0.5 \times 0.5 \times 0.5$ mm$^3$ |
| **Alzheimer's Disease** | 50 (28) | 75 ± 7.3 | ToF | $0.3 \times 0.3 \times 0.3$ mm$^3$ |

Segmentation and Registration

We used our recently developed methodology to characterize vascular morphology from the imaging data. [18]. Using automatic cerebral vascular segmentation, we extracted vessels as fine as the imaging resolution to determine the geometric properties of the segmented vascular network. To provide a common standard space for comparison of all the scans in the study and account for the varying reference coordinates between different scanners and patient positions, each time-of-flight (ToF) MRA sequence and the corresponding segmentation were spatially normalized to the

Montreal Neurological Institute (MNI) standard anatomical brain atlas reference space [25] with 0.5 mm$^3$ isotropic image resolution in three steps: first, each patient's ToF MRA sequences were registered to the corresponding T1-weighted imaging (T1) dataset via a rigid transformation. The T1 datasets were then registered to the T1 MNI brain atlas through an affine transformation. Finally, the two transformations were concatenated and multiplied to transform the segmented network directly into the reference space of the MNI brain atlas. The MNI atlas, adopted by the International Consortium of Brain Mapping, uses a large series of 152 normal control MRI scans to define standard anatomy and consists of 'templates' for MRI sequences routinely used in both clinical and research settings to register MR data to common space.

Healthy cerebrovascular atlas and geometric feature analysis
We averaged the registered segmentations from the 175 healthy ToF scans to create the atlas of regional artery probability, with every pixel containing the probability of belonging to a vessel. We then extracted the average geometric features of the healthy brain vessel networks including vessel tortuosity, fractal dimension, number of branches, average and total branch length, and diameter [18]. These properties have been used to study the associated changes in vascular stiffness and tortuosity within healthy adults [18,26,27] and are corroborated indicators of vascular health and disease [8,18–20]. The anatomical regions in the atlas were labeled based on major vascular territories, namely the left and right middle cerebral artery (MCA), the anterior cerebral artery (ACA) and the posterior cerebral artery (PCA). This labeled cerebrovascular atlas of the healthy population, including the averaged geometric features of the vessel network, was used to study the variations in CoW in healthy subjects and vessel geometry due to aging. To quantify aging-related changes in the vasculature, the healthy cohort was divided into five age groups: group 1 (< 29-year-old, n=22), group 2 (30-39-year-old, n=20), group 3 (40-49-year-old, n=21), group 4 (50-59-year-old, n=19), and group 5 (> 60-year-old, n=17). Given its key role in maintaining CBF distribution and providing collateral blood flow, CoW variations can affect the clinical outcomes after a cerebrovascular event. We assessed CoW variations within these networks, including a fully formed CoW and an absent or hypoplastic left and right anterior and posterior communicating arteries (ACoA and PCoA, respectively) and the first segment of the PCA (P1).

Quantifying alterations in vascular geometry in AIS patients
Emergent large vessel occlusion (ELVO) strokes occur upon the occlusion of a major intracranial artery (ICA, M1, M2 segments of the MCA and Basilar) and significantly decrease CBF to large areas of the brain. Patients with ELVO stroke are expected to show major deviations from the healthy atlas with a significant decrease in the number of branches, total length, and volume of the cerebral vasculature. Nevertheless, other geometric variations such as vessel tortuosity and fractality have not been studied yet. The stroke data consisted of 45 AIS patients with confirmed clinical and radiographic diagnosis of occlusion of the first (M1) and second (M2) segments of the MCA (n=24 and n=9, respectively) and the internal carotid artery (ICA, n=12), corresponding to the most common stroke syndromes in the average population [28]. We defined a "vaso-deviation score" by determining how different each voxel of a patient's specific vasculature is from its healthy counterpart in the atlas and normalizing this difference by the variance within the atlas. Regions with a higher vaso-deviation score correspond to the ELVO location.

Quantifying alterations in vascular geometry in AD patients
We hypothesized that the vasculature of AD patients possesses abnormally high branching and complexity with a higher tortuosity, fractal dimensions and number of branches, and a decreased average vessel diameter. We applied our validated segmentation and feature extraction method to MRA scans of 50 AD patients with varying levels of dementia, defined using the CDR scale [24] and extracted the cerebral vascular networks and their corresponding geometric properties. The quantitative extracted geometric features were compared to the healthy average obtained from the atlas. The data was then further divided into two - early-mild (CDR 0.5 and 1) and moderate-severe (CDR 2 and 3) dementia levels [24]. We tested for significant quantitative differences in the vasculature between AD patients versus healthy subjects as well as between patients with early-mild versus moderate-severe dementia.

Statistical analysis
When applicable, we conducted one-way ANOVA, followed by Tukey's post hoc test to assess changes in the vascular architecture across the age groups within healthy subjects. The comparison of stroke and healthy vasculature was performed for 46 stroke patients, which were quantitatively compared against the healthy subjects from the atlas using the ANOVA test, after adjusting for age as a factor. The vascular networks of age-matched healthy controls and AD subjects were compared using the multi-way ANOVA between groups (healthy, early-mild AD, moderate-severe AD). To avoid inconsistency in the data for quantitative analysis, we only compared data with the same imaging resolution.

**Results**

Healthy cerebrovascular atlas
The generated probabilistic cerebrovascular atlas displayed the major cerebral arteries based on occurrence in specific anatomical locations. The results were visually and quantitatively in accordance with the expected occurrence probabilities of regular cerebrovascular anatomy. The largest probabilities of occurrence were seen at the base of the ICA (immediately after bifurcation from CCA), followed by the ICA bifurcations into bilateral MCAs and ACAs, and then the BA. Compared to the anterior circulation, the higher variation in the BA diameter in healthy adults resulted in scattered raw probability values. Figure 1 shows three axial slices of arterial probabilities overlaid on the corresponding axial T1 slices. The slices have been chosen from different axial locations as shown in the corresponding panel (purple inserts) ranging from the base of the CoW to superior regions of the brain. The color bar contains the raw values from the atlas map and represents voxels with varying probability of being on a vessel. With an expected reduction in vessel diameters and increased variation among individuals, the arterial occurrence probability decreased upon moving distally from the CoW.

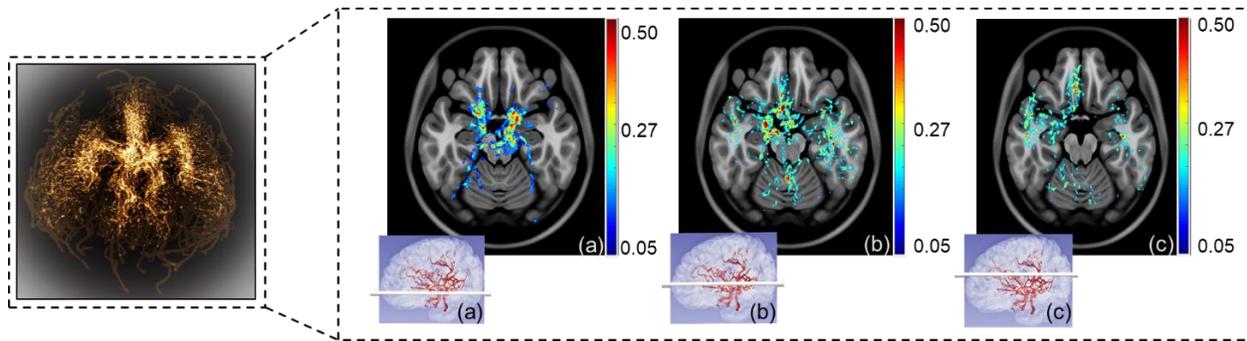

**Figure 1.** The probabilistic atlas – selected axial slices showing the varying probability of arterial occurrence. The left panel shows the top view of the 3D atlas with brighter regions corresponding to higher probability values.

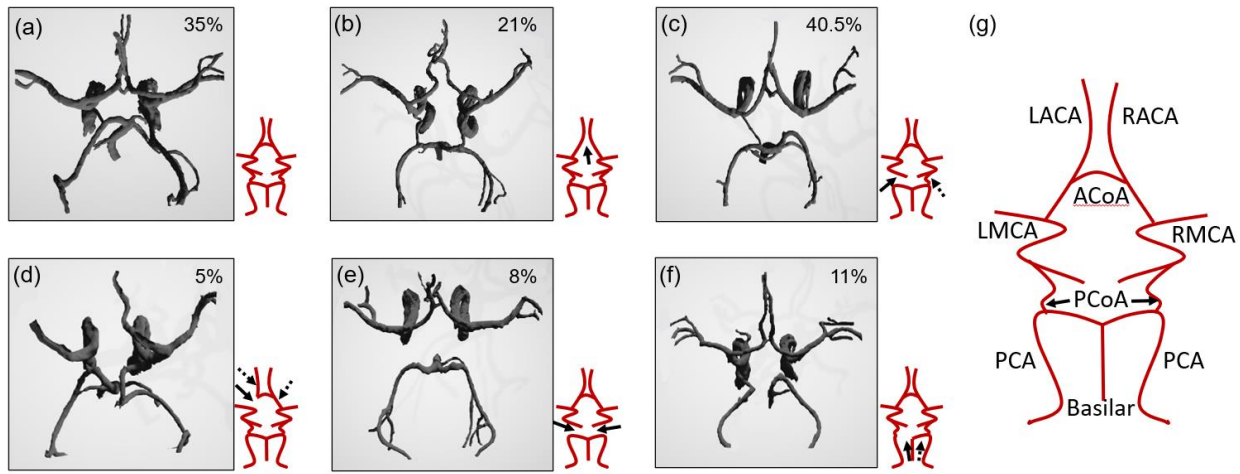

**Figure 2.** The 6 most commonly occurring variations in the CoW anatomy of our healthy dataset. (a) Fully connected and formed CoW, (b) ACoA missing, (c) one of the PCoA missing, (d) A1 segment of ACA missing, (e) both of the PCoA missing, (f) P1 segment of PCA from Basilar missing. The arrows depict the missing segments (dashed arrows imply the possibility of the other equivalent segment being missing as well as the one shown by the solid arrow) and (g) a labeled schematic representation of the fully formed CoW, showing all the major vessels. The percentage values in each panel corresponds to the frequency of occurrence of that model within the healthy population used to study the variations.

Table 2 summarizes the geometric features extracted from 175 healthy subjects included in the statistical atlas. We found that only 35% of healthy adults had a fully formed CoW, whereas 40.5% missed at least one PCoA, 21% missed the ACoA, and 11% and 8% missed the A1 and P1 segments of the ACA and PCA, respectively. A smaller number of healthy adults missed one ACA (1%), P1 segment of the PCA (2%) or the PCoA (5%). We include the six most common presentations of the CoW anatomy within our dataset in Figure 2. The variations are found to be generated due to under-developed or missing segments in most cases.

Figure 3 includes the major findings of the effects of aging on the healthy vascular variations. There was a significant increase across most age groups in the tortuosity ($p<0.001$), total length ($p<0.001$), number of branches ($p<0.001$), and fractal dimension ($p<0.001$) in a multi-group

ANOVA. Post hoc analysis showed a significant increase in tortuosity and fractality with age which is consistent with our expectations. There was a significant increase in total length with age which would be expected since there was also an increase in the number of branches, which would increase the overall length of the vasculature. No significant overall change was reported in average diameter, average branch length or maximum branch length across the 5 age groups.

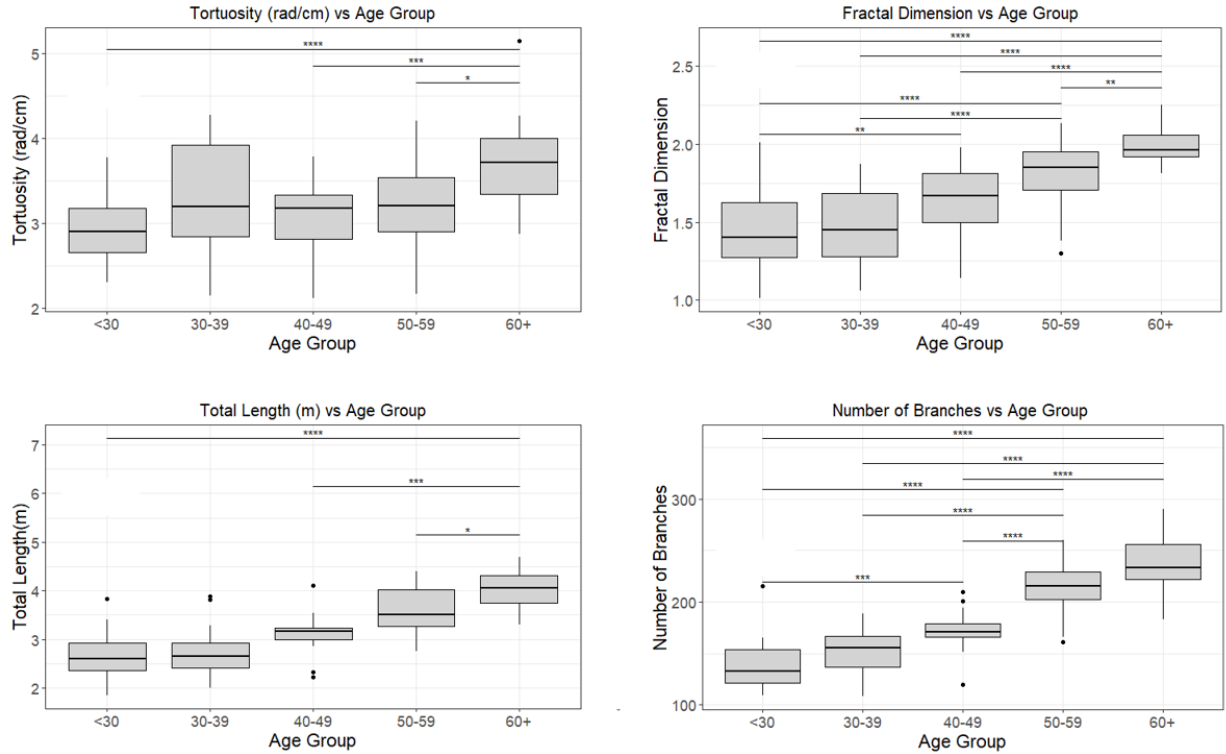

**Figure 3.** Major findings of the effect of aging conducted on the healthy subjects from the atlas, demonstrating a significant change in vascular patterns with increasing age in certain vascular features. The ANOVA test results between age groups for the significantly different groups are represented by * for p<0.05, ** for p<0.01 and *** for p<0.001. The multi-group ANOVA had p<0.001 for tortuosity, number of branches and fractal dimension.

**Table 2.** A comparison of geometric features of the cerebral vascular tree of healthy subjects and Stroke patients (mean +/- SD). The significantly different features are highlighted in **bold**.

|  | Healthy subjects | Stroke patients | *p* - value |
| --- | --- | --- | --- |
| Total length (m) | 3.19 ± 0.67 | 2.10 ± 0.71 | **0.005** |
| Number of branches | 179 ± 43.76 | 166 ± 75.69 | 0.051 |
| Average branch length (mm) | 18.77 ± 4.00 | 12.58 ± 2.07 | **<0.001** |
| Maximum branch length (mm) | 116.50 ± 25.22 | 59.38 ± 6.10 | 0.738 |
| Average diameter (mm) | 2.75 ± 0.51 | 2.18 ± 0.38 | **0.007** |
| Total volume (ml) | 96.12 ± 17.51 | 63.45 ± 21.83 | **0.013** |
| Fractal dimension | 1.55 ± 0.29 | 1.79 ± 0.20 | **0.007** |
| Tortuosity (rad/cm) | 3.24 ± 0.57 | 5.80 ± 0.92 | **<0.001** |

Vascular geometry in stroke patients

The geometric features of the cerebral vascular tree of 45 stroke patients compared to the healthy atlas of subjects and the average values for each geometric feature are presented in Table 2. As supported by our previously published preliminary findings, cerebral vasculature is significantly altered in stroke patients, beyond the expected shorter total length (2.10±0.71 m vs. 3.19±0.67 m) and smaller volume (63.45±21.83 ml vs. 96.12±17.51 ml). Stroke brains were also found to possess higher tortuosity (5.80±0.92 vs. 3.24±0.57) and fractal dimension (1.79±0.2 vs. 1.55±0.29), alluding to atypical vessel remodeling, along with the expected findings of reduced number of branches, length, and volume. These features show tremendous potential to be used as a tool for automatic detection of occlusion in stroke patients in the clinic.

Vascular geometry in AD patients

As hypothesized, the AD patients' vascular network is characterized by higher tortuosity and fractality, greater number of branches and a smaller average diameter. These effects of disease were seen throughout the AD group, when adjusted for age. As dementia levels increase from early-mild to moderate-severe, we observed significant differences in the vascular features between groups with varying levels of dementia, specifically higher tortuosity fractality, number of branches, and a smaller average diameter. The quantitative information of the vascular geometric features for 50 patients with AD have been reported in Table 3, including a comparison with healthy subjects. For visual assessment, Figure 4 shows the cerebral vascular network from 2 healthy subjects and 2 AD patients and depicts the heightened tortuosity including looping and twisting in the vascular network of an AD patient. Such twists and loops are not observed in healthy subjects and is corroborated with studies which reported similar behavior of cerebral vessels in mice (6). Figure 5 shows plots of the significantly different features between the healthy and AD groups, namely the total length, fractal dimension, tortuosity, and average diameter.

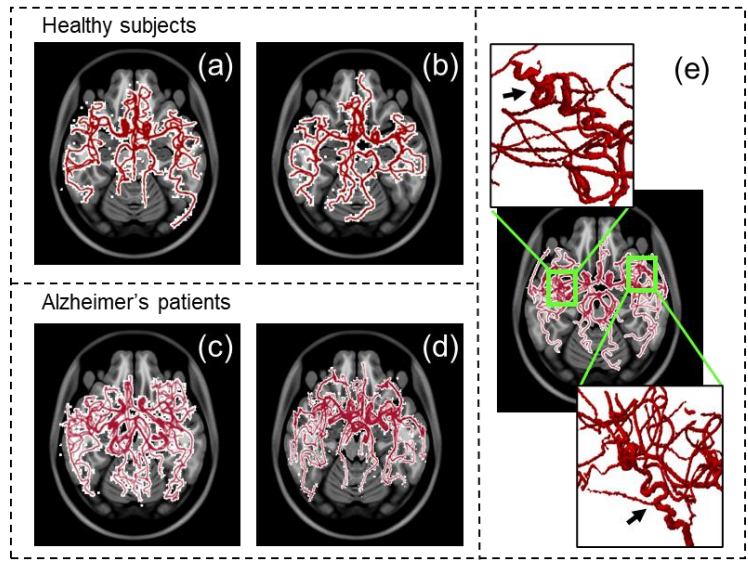

**Figure 4.** Cerebral vascular network from 2 healthy subjects (a) and (b) and 2 AD patients (c) and (d) for visual comparison. In panel e, we can see the elevated tortuosity including looping and twisting in the vascular network of an AD patient.

**Table 3.** A comparison of geometric features of the cerebral vascular tree of healthy subjects and AD patients (mean +/- SD)

| | Healthy subjects | AD patients (CDR = 0.5, 1) | AD patients (CDR = 2) | *p* - value |
|---|---|---|---|---|
| Total length (m) | 9.27 ± 3.10 | 11.01 ± 2.87 | 14.35 ± 3.36 | **0.013** |
| Number of branches | 280 ± 140 | 299.1 ± 136 | 413.87 ± 104.20 | 0.096 |
| Avg diameter (mm) | 2.75 ± 0.14 | 2.59 ± 0.18 | 2.31 ± 0.11 | **0.001** |
| Fractal dimension | 1.83 ± 0.11 | 1.89 ± 0.03 | 1.98 ± 0.05 | **0.001** |
| Tortuosity (rad/cm) | 7.98 ± 3.71 | 10.09 ± 2.46 | 12.13 ± 3.17 | **0.008** |

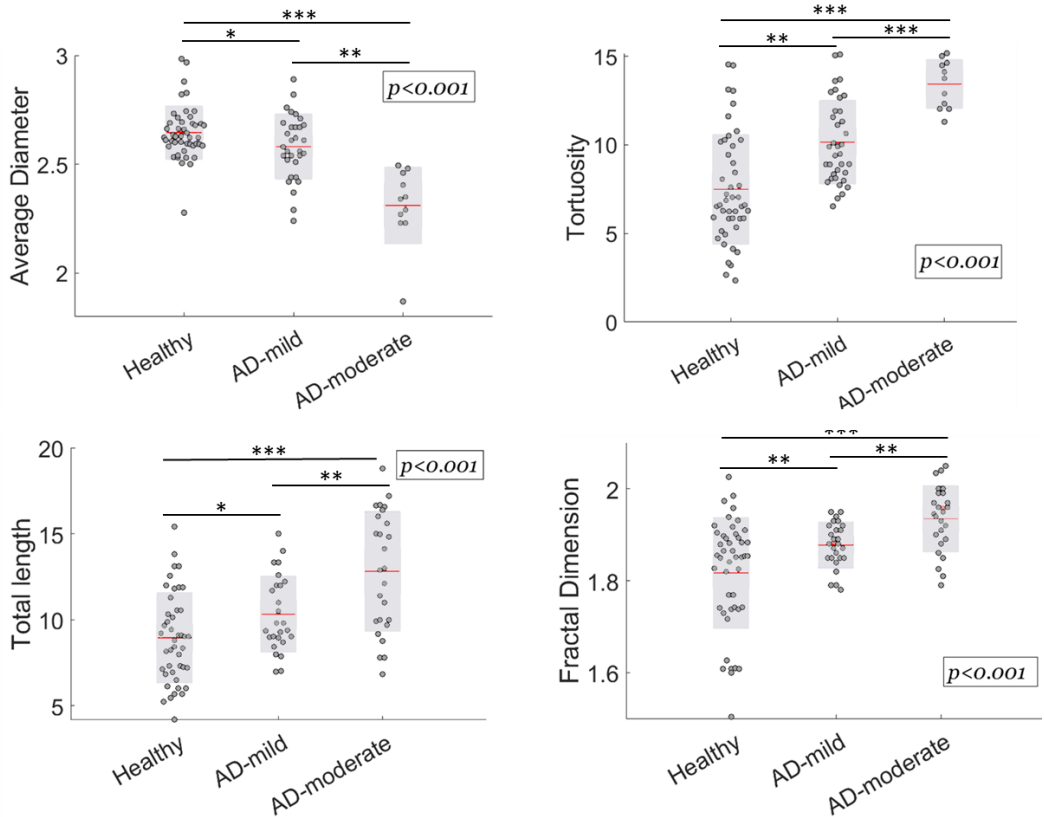

**Figure 5.** The scatter plots of vascular geometrical features of the healthy subjects, early-mild, and moderate-severe Alzheimer's disease (AD). The red horizontal lines and shaded regions represent the mean and one standard deviation, respectively.

**Discussion**

Both natural aging and disease alter the cerebrovascular network. These changes, from the cellular level to vessel wall deterioration and ultimately modified vascular architecture, contribute to the development and progression of neurologic disorders [13,14]. Therefore, a thorough understanding of cerebrovascular alterations such as increase in arterial stiffness, tortuosity, endothelial dysfunction and hypoperfusion, offers insight into the underlying pathophysiological pathways, natural course, and risk assessment of common and burdensome cerebrovascular and neurodegenerative diseases. A quantitative analysis of the cerebrovascular network may improve the timing and accuracy of diagnosis and provide an opportunity for preventive and early therapeutic interventions for individuals at risk for ischemic and hemorrhagic strokes, atherosclerosis and dementia. The analysis of modified vascular structure with the natural aging process offers insight into the various pathways leading to these cerebrovascular alterations and possibly causing or exacerbating disease in older adults [17]. The comprehensive atlas of healthy cerebral vasculature enables quantitative characterization of the average "normal" population based on vascular geometry features. The healthy template for vessel distribution and geometrical patterns can enable clinicians and researchers to compare and analyze changes in the cerebrovascular network during aging and various pathological states.

We found a significant age-related increase in vessel tortuosity and fractality that can be attributed to the increased endothelial resistance, and in turn, an increased incidence of hypertension in older adults. An unexpected finding was the lack of significant change in average diameter, contrary to a previous report of increased vessel diameter by aging [27]. Additional studies including a larger sample size across all age groups, particularly older than 70, are warranted to elucidate this finding with implications in risk-stratification for cerebrovascular disorders.

The CoW plays a pivotal role in CBF distribution in both healthy and pathological conditions. For instance, the CoW anatomy, by governing collateral blood flow during AIS, can alter the natural course of ischemic core growth, salvageable penumbra, response to acute reperfusion therapies, and ultimately clinical outcomes [29,30]. However, the CoW anatomy varies considerably among individuals [12,31]. Therefore, we found it imperative to quantify various anatomic presentations of the CoW in healthy adults. Using a large database of healthy subjects, we found that only 35% of individuals have a completely formed CoW, consistent with earlier reports, while the rest of the population present some anatomical variation in CoW anatomy [12,31]. This finding is primarily due to hypoplastic (underdeveloped) or aplastic (absent) rather than fused or additional vessel segments.

Using the labeled healthy cerebrovascular atlas, we quantified the deviation in the vascular networks of stroke patients from the healthy population. Similar to our previous preliminary report [18], We found a marked change in cerebrovascular geometry of stroke patients consisting of an expected decrease in total vessel length, volume, and number of branches along with an increased vessel tortuosity and fractality. These findings can be attributed to atypical vessel morphology in stroke patients. The deviations from healthy cerebrovascular geometry can serve as a diagnostic tool for automated stroke detection to complement the existing clinical and imaging capabilities [32]. Such a tool has several implications in both bedside and bench assessment and treatment of stroke patients. Analysis of vascular geometric features can enhance timing, sensitivity, and specificity of occlusion detection not only for proximal large vessels but also for distal medium vessels that may also be amenable to acute reperfusion treatments. These advantages are particularly important when there is a clinical equipoise or lack of access to subspecialized neuroradiologists to interpret multimodal advanced neuroimaging studies. Moreover, assessment of vascular morphology can improve the prediction of stroke patients' response to reperfusion therapies [9] and long-term clinical outcomes [33] since recent technological advances have expanded the application of endovascular thrombectomy as an increasingly effective treatment for AIS [34]. Interestingly, the altered vessel morphology may precede the ischemic event in stroke patients, alluding to its possible application as an early marker for stroke occurrence and prevention, but this needs to be studied extensively in longitudinal studies with a large dataset.

Neurodegenerative diseases, including AD, are characterized by irreversible and progressive neuronal loss. AD is increasingly recognized as a multifactorial disease with well-established vascular risk factors [35] and a key role for vascular dysfunction [4,7,36,37]. Neuritic plaques and vascular amyloid depositions, the early hallmarks of AD are associated with altered cerebral vasculature. Documented hypoperfusion in AD can initiate a vicious cycle of amyloid deposition, neurodegeneration, and cognitive decline [2,14,38]. Accordingly, vascular morphological changes have been reported in AD patients, with some changes preceding clinical presentations. Using non-invasive MRA data, we performed a quantitative analysis of vascular geometric changes in AD patients. We found higher vessel tortuosity and fractality in AD patients, compared to age-adjusted healthy subjects, further enhanced by increasing levels of dementia. These findings agree with the previously hypothesized correlation between vessel stiffness and resistance, seen in AD patients, and structural vessel changes [36,39]. The observed higher number of branches in AD patients can possibly be due to the known abnormal vessel branching and angiogenesis induced by amyloid plaque deposition in vessel walls. The progressively heightened vascular complexity correlated with disease progression suggests that vascular geometry analysis of ToF MRA can be used as a fast, automated, and non-invasive early biomarker as well as a surveillance tool for disease progression in AD and possibly other neurodegenerative diseases.

Limitations

Our study has a few limitations. The extracted geometric properties are inherently affected by the source imaging resolution and region of interest (ROI). The algorithm identifies vessels with a diameter as small as the imaging resolution; therefore, a higher resolution increases the number of smaller vessels detected. The difference in spatial resolution of MIDAS ($0.5 \times 0.5 \times 0.5$ mm$^3$) and OASIS-3 ($0.27 \times 0.27 \times 0.27$ mm$^3$) datasets may explain the lower average values for total length, volume, and the number of branches for healthy adults derived from MIDAS (see Tables 2 and 3 for comparison). In addition, the MRA scans of most of the subjects in the OASIS-3 dataset had a larger ROI due to the image acquisition starting at a lower axial slice and including the vertebral arteries and a larger portion of the ICA. To avoid any discrepancy in the data and maintain the quantitative analysis' cogency, we restricted the comparison to data with the same imaging resolution. The features obtained from the stroke scans (also imaged at $0.5 \times 0.5 \times 0.5$ mm$^3$) were only compared to those of healthy subjects from the MIDAS data, and the features of AD subjects were compared only within the OASIS-3 dataset. We used MRA data of 45 patients with MCA and ICA occlusions to analyze cerebrovascular morphological changes in stroke patients. Although this sample represents the most common large vessel occlusion sites in AIS [40], a larger sample size, including other proximal large and distal medium vessel occlusion sites, can improve our understanding of vascular architecture changes in stroke.

Conclusion

In conclusion, we developed a comprehensive atlas of healthy cerebrovascular morphology. Compared to healthy individuals, stroke and AD patients showed a significantly altered vascular morphology. The probabilistic atlas has major implications as a non-invasive automated tool to study healthy aging and various pathological states, including prediction, diagnosis, surveillance, and treating cerebrovascular and neurodegenerative diseases. Further studies, including a larger sample size of various neurologic diseases and corresponding clinical data, are warranted to establish future clinical applications of the algorithm and atlas.

**Acknowledgements and Disclosure:**

This study was supported by an award by NIH NINDS (National Institute of Neurological Disorders and Stroke), grant number R03NS108167.The findings of this manuscript are those of the authors and do not necessarily represent the official views of NINDS. The authors declare that they have no competing interests.

**References**

1. World Health Organization. World Health Statistics. 2019.

2. Kalaria RN. Cerebral vessels in ageing and Alzheimer's disease. *Pharmacol Ther*. 1996;72(3):193-214. doi:10.1016/S0163-7258(96)00116-7

3. Wei F, Diedrich KT, Fullerton HJ, et al. Arterial Tortuosity: An Imaging Biomarker of Childhood Stroke Pathogenesis? *Stroke*. 2016;47(5):1265-1270. doi:10.1161/STROKEAHA.115.011331

4. Arvanitakis Z, Capuano AW, Leurgans SE, Bennett DA, Schneider JA. Relation of cerebral vessel disease to Alzheimer's disease dementia and cognitive function in elderly people: a cross-sectional study. *Lancet Neurol*. 2016;15(9):934-943. doi:10.1016/S1474-4422(16)30029-1

5. Meyer EP, Ulmann-Schuler A, Staufenbiel M, Krucker T. Altered morphology and 3D architecture of brain vasculature in a mouse model for Alzheimer's disease. *Proc Natl Acad Sci U S A*. 2008;105(9):3587-3592. doi:10.1073/pnas.0709788105

6. Amukotuwa SA, Straka M, Smith H, et al. Automated detection of intracranial large vessel occlusions on computed tomography angiography a single center experience. *Stroke*. 2019;50(10):2790-2798. doi:10.1161/STROKEAHA.119.026259

7. Sweeney MD, Montagne A, Sagare AP, et al. Vascular dysfunction—The disregarded partner of Alzheimer's disease. *Alzheimer's Dement*. 2019;15(1):158-167. doi:10.1016/j.jalz.2018.07.222

8. Wright SN, Kochunov P, Mut F, et al. Digital reconstruction and morphometric analysis of human brain arterial vasculature from magnetic resonance angiography. *Neuroimage*. 2013;82:170-181. doi:10.1016/j.neuroimage.2013.05.089

9. Liu J, Wang Y, Akamatsu Y, et al. Vascular remodeling after ischemic stroke: Mechanisms and therapeutic potentials. *Prog Neurobiol*. 2014;115(C):138-156. doi:10.1016/j.pneurobio.2013.11.004

10. Tanaka M, Sakaguchi M, Miwa K, et al. Basilar artery diameter is an independent

predictor of incident cardiovascular events. *Arterioscler Thromb Vasc Biol*. 2013;33(9):2240-2244. doi:10.1161/ATVBAHA.113.301467

11. Mukherjee D, Jani ND, Narvid J, Shadden SC. The Role of Circle of Willis Anatomy Variations in Cardio-embolic Stroke: A Patient-Specific Simulation Based Study. *Ann Biomed Eng*. 2018;46(8):1128-1145. doi:10.1007/s10439-018-2027-5

12. Alastruey J, Parker KH, Peiro J, Byrd SM, Sherwin SJ. Modelling the circle of Willis to assess the effects of anatomical variations and occlusions on cerebral flows. *J Biomech*. 2007;40(8):1794-1805. doi:10.1016/j.jbiomech.2006.07.008

13. Kalaria RN, Akinyemi R, Ihara M. Does vascular pathology contribute to Alzheimer changes? *J Neurol Sci*. 2012;322(1-2):141-147. doi:10.1016/j.jns.2012.07.032

14. Toledo JB, Arnold SE, Raible K, et al. Contribution of cerebrovascular disease in autopsy confirmed neurodegenerative disease cases in the National Alzheimer's Coordinating Centre. *Brain*. 2013;136(9):2697-2706. doi:10.1093/brain/awt188

15. Fischer VW, Siddiqi A, Yusufaly Y. Altered angioarchitecture in seleced areas of brains with Alzheimer's disease. *Acta Neuropathol*. 1990;79:672-679.

16. Barbará-Morales E, Pérez-González J, Rojas-Saavedra KC, Medina-Bañuelos V. Evaluation of Brain Tortuosity Measurement for the Automatic Multimodal Classification of Subjects with Alzheimer's Disease. *Comput Intell Neurosci*. 2020;2020. doi:10.1155/2020/4041832

17. Donato AJ, Machin DR, Lesniewski LA. Mechanisms of dysfunction in the aging vasculature and role in age-related disease. *Circ Res*. 2018;123(7):825-848. doi:10.1161/CIRCRESAHA.118.312563

18. Deshpande A, Jamilpour N, Jiang B, et al. Automatic segmentation, feature extraction and comparison of healthy and stroke cerebral vasculature. *NeuroImage Clin*. 2021;30(January):102573. doi:10.1016/j.nicl.2021.102573

19. Chen L, Mossa-Basha M, Balu N, et al. Development of a quantitative intracranial vascular features extraction tool on 3D MRA using semiautomated open-curve active contour vessel tracing. *Magn Reson Med*. 2018;79(6):3229-3238. doi:10.1002/mrm.26961

20. Kim BJ, Kim SM, Kang DW, Kwon SU, Suh DC, Kim JS. Vascular tortuosity may be related to intracranial artery atherosclerosis. *Int J Stroke*. 2015;10(7):1081-1086. doi:10.1111/ijs.12525

21. Cool D, Chillet D, Kim J, Guyon JP, Foskey M, Aylward S. Tissue-based affine registration of brain images to form a vascular density atlas. *Lect Notes Comput Sci*. 2003;2879(PART 2):9-15. doi:10.1007/978-3-540-39903-2_2

22. Mouches P, Forkert ND. A statistical atlas of cerebral arteries generated using multi-center MRA datasets from healthy subjects. *Sci data*. 2019;6(1):29. doi:10.1038/s41597-019-0034-5

23. Zanto TP, Hennigan K, Östberg M, Clapp WC, Gazzaley A. Vessel Tortuosity and Brain Tumor Malignancy: A Blinded Study. 2011;46(4):564-574.

doi:10.1016/j.cortex.2009.08.003.Predictive

24. LaMontagne P. Longitudinal Neuroimaging, Clinical, and Cognitive Dataset for Normal Aging and Alzheimer Disease. *arXiv Prepr arXiv181207731*. 2019.

25. Grabner G, Janke AL, Budge MM, Smith D, Pruessner J, Collins DL. Symmetric atlasing and model based segmentation: An application to the hippocampus in older adults. *Lect Notes Comput Sci (including Subser Lect Notes Artif Intell Lect Notes Bioinformatics)*. 2006;4191 LNCS:58-66. doi:10.1007/11866763_8

26. Xu X, Wang B, Ren C, et al. Age-related Impairment of Vascular Structure and Functions. 2017;8(5):590-610.

27. Bullitt E, Zeng D, Mortamet B, et al. The effects of healthy aging on intracerebral blood vessels visualized by magnetic resonance angiography. *Neurobiol Aging*. 2010;31(2):290-300. doi:10.1016/j.neurobiolaging.2008.03.022

28. Blood Vessels of the Brain | Internet Stroke Center. http://www.strokecenter.org/professionals/brain-anatomy/blood-vessels-of-the-brain/. Accessed March 6, 2020.

29. Gregory W Albers1, Maarten G Lansberg1, Stephanie Kemp1, Jenny P Tsai1, Phil Lavori1, Soren Christensen1, Michael Mlynash1, Sun Kim1, Scott Hamilton1, Sharon D Yeatts2, Yuko Palesch2, Roland Bammer1, Joe Broderick3 and MPM. A multicenter randomized controlled trial of endovascular therapy following imaging evaluation for ischemic stroke (DEFUSE 3). *Int J stroke*. 2017;176(3):139-148. doi:10.1177/1747493017701147.A

30. Stryker Neurovascular. Clinical Mismatch in the Triage of Wake Up and Late Presenting Strokes Undergoing Neurointervention With Trevo (DAWN). *Stroke*. 2018;49(2):498-500. doi:10.1161/STROKEAHA.117.018560

31. Ren Y, Chen Q, Li ZY. A 3D numerical study of the collateral capacity of the Circle of Willis with anatomical variation in the posterior circulation. *Biomed Eng Online*. 2015;14(Suppl 1):S11. doi:10.1186/1475-925X-14-S1-S11

32. Sarmento RM, Vasconcelos FFX, Filho PPR, Wu W, De Albuquerque VHC. Automatic Neuroimage Processing and Analysis in Stroke - A Systematic Review. *IEEE Rev Biomed Eng*. 2020;13:130-155. doi:10.1109/RBME.2019.2934500

33. Alaka SA, Menon BK, Brobbey A, et al. Functional Outcome Prediction in Ischemic Stroke: A Comparison of Machine Learning Algorithms and Regression Models. *Front Neurol*. 2020;11(August):889. doi:10.3389/fneur.2020.00889

34. Saver JL, Chapot R, Agid R, et al. Thrombectomy for Distal, Medium Vessel Occlusions: A Consensus Statement on Present Knowledge and Promising Directions. *Stroke*. 2020;(September):2872-2884. doi:10.1161/STROKEAHA.120.028956

35. Santos CY, Snyder PJ, Wu WC, Zhang M, Echeverria A, Alber J. Pathophysiologic relationship between Alzheimer's disease, cerebrovascular disease, and cardiovascular risk: A review and synthesis. *Alzheimer's Dement Diagnosis, Assess Dis Monit*. 2017;7:69-87. doi:10.1016/j.dadm.2017.01.005

36. Govindpani K, McNamara LG, Smith NR, et al. Vascular Dysfunction in Alzheimer's Disease: A Prelude to the Pathological Process or a Consequence of It? *J Clin Med*. 2019;8(5):651. doi:10.3390/jcm8050651

37. Diagnosis and Treatment of Clinical Alzheimer's-Type Dementia. *Eff Heal Care Progr*. 2020.

38. Klohs J, Rudin M, Shimshek DR, Beckmann N. Imaging of cerebrovascular pathology in animal models of Alzheimer's disease. *Front Aging Neurosci*. 2014;6(MAR). doi:10.3389/fnagi.2014.00032

39. El Tannir El Tayara N, Delatour B, Volk A, Dhenain M. Detection of vascular alterations by in vivo magnetic resonance angiography and histology in APP/PS1 mouse model of Alzheimer's disease. *Magn Reson Mater Physics, Biol Med*. 2010;23(1):53-64. doi:10.1007/s10334-009-0194-y

40. Saver JL, Goyal M, Van Der Lugt A, et al. Time to treatment with endovascular thrombectomy and outcomes from ischemic stroke: Ameta-analysis. *JAMA - J Am Med Assoc*. 2016;316(12):1279-1288. doi:10.1001/jama.2016.13647